\documentclass[journal,10pt]{IEEEtran}
\usepackage{booktabs}
\usepackage{caption,subcaption}

\ifCLASSINFOpdf
  \usepackage[pdftex]{graphicx}
\else
\fi
\usepackage{color}

\usepackage[cmex10]{amsmath}
\usepackage{amssymb}
\usepackage{algorithmic}

\usepackage{stfloats}

\usepackage{threeparttable}

\begin{document}
%
\title{Hardware Architecture for List Successive Cancellation Decoding of Polar Codes}
%
%
%

\author{Alexios~Balatsoukas-Stimming,~\IEEEmembership{Student~Member,~IEEE}, Alexandre~J.~Raymond,\\Warren~J.~Gross,~\IEEEmembership{Senior Member,~IEEE}~and~Andreas~Burg,~\IEEEmembership{Member,~IEEE}}
\maketitle

\begin{abstract}
We present a hardware architecture and algorithmic improvements for list SC decoding of polar codes. More specifically, we show how to completely avoid copying of the likelihoods, which is algorithmically the most cumbersome part of list SC decoding. The hardware architecture was synthesized for a blocklength of $N = 1024$ bits and list sizes $L=2,4$ using a UMC~90nm VLSI technology. The resulting decoder can achieve a coded throughput of $181$~Mbps at a frequency of $459$~MHz.

\end{abstract}


%
\IEEEpeerreviewmaketitle

\section{Introduction}
%
%
%
%
\IEEEPARstart{C}{hannel} polarization gives rise to an elegant and provably good class of channel codes, called \emph{polar codes}~\cite{Arikan2009}. 
Decoding of polar codes is usually performed using a successive cancellation (SC) decoder~\cite{Arikan2009}. Some hardware architectures for SC decoding of polar codes were discussed in~\cite{Leroux2011,Leroux2013,Zhang2012}, and \cite{Zhang2013}, while the first ASIC of such a decoder was presented in~\cite{Mishra2012}. Moreover, an FPGA implementation of a belief propagation decoder for polar codes was presented in \cite{Pamuk2011}. Recently, more sophisticated decoding algorithms, such as the list SC decoder \cite{Tal2011,Chen2013}, and the stack SC decoder \cite{Chen2013}, were introduced. These algorithms provide improved error correcting performance at the cost of increased complexity. Stack SC decoding suffers from high memory requirements, costly metric normalization, and non-deterministic decoding latency, making list SC decoding more attractive from a practical perspective. Unfortunately, the list SC decoder is burdened by a likelihood copying step and no architecture of such a decoder exists yet in the literature.


\subsubsection*{Contribution and Outline} This brief presents an architecture for list SC decoding of polar codes. To this end, we also describe how the copying of the intermediate likelihoods in the list SC decoding algorithm can be avoided. In Section \ref{sec:polardec}, we briefly review the construction and decoding of polar codes. Section~\ref{sec:alg} discusses algorithmic improvements to list SC decoding, while in Section~\ref{sec:arch} the proposed list SC decoder architecture is described. Section~\ref{sec:results} summarizes VLSI implementation results and  concludes this letter.

\section{Polar Codes}\label{sec:polardec}


We use $a_1^N$ to denote a row vector $(a_1,\hdots,a_N)$ and $a_i^j$ to denote the subvector $(a_i,\hdots,a_j)$. We use the operators $\log$ and $\ln$ for the binary and natural logarithm, respectively.

A polar code is constructed by recursively applying a polarizing transform $n$ times to the binary input symmetric and memoryless channel $W$. This transform is linear and it can be expressed as a $2 \times 2$ matrix, denoted by $\mathbf{F}$. The $n$-fold application of this transform can be expressed as an $N \times N$ matrix $\mathbf{G}$, with $\mathbf{G} = \mathbf{F}^{\otimes n}$, where $^{\otimes n}$ denotes the $n$-fold application of the Kronecker product. Encoding is performed by choosing a sequence $u_1^N \in \{0,1\}^N$ and calculating the codeword $x_1^N = u_1^N\mathbf{G}$. This codeword is transmitted over $N$ uses of $W$ and a noisy codeword $y_1^N$ is received.

\subsection{Successive Cancellation Decoding}

The decoding method proposed by Ar{\i}kan is based on successive cancellation. First, an estimate for $u_1$, denoted by $\hat{u}_1$, is calculated based on $y_1^N$. Then, $u_2$ is decoded, based on $y_1^N$ \emph{and} the knowledge of $\hat{u}_1$, etc. In principle, it is possible to calculate the mutual information between $(y_1^N,u_1^{i-1})$ and $u_i$ for every $i$. A polar code of rate $R$ is constructed by letting only the $NR$ $u_i$'s with the highest mutual information convey information, while freezing the remaining $u_i$'s to $0$. The sets of non-frozen and frozen bit indices are denoted by $\mathcal{A}$ and $\mathcal{A}^c$, respectively.
\begin{figure}
	\centering
	\includegraphics[width=0.30\textwidth]{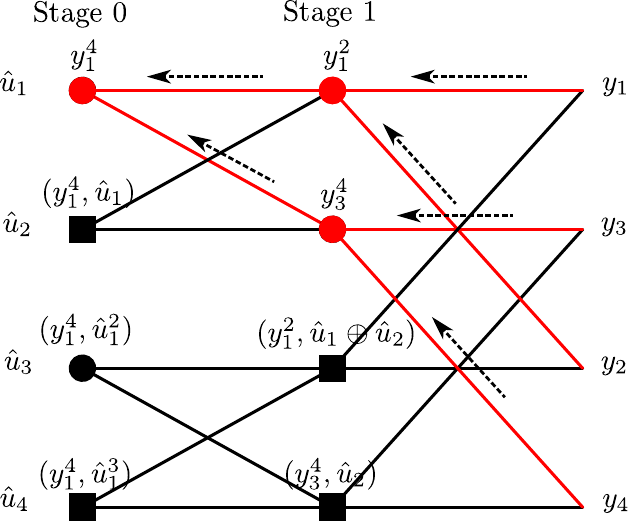}
	\caption{SC decoding of $u_1$ with $N=4$.}
	\label{fig:sc}
\end{figure}
The exact decoding procedure is dictated by the recursive structure of the code. In Fig.~\ref{fig:sc}, the decoding process is visualized for $N=4$. On the right-hand side of the graph, the likelihoods $W(y_i|x_i),~{i = 1,\hdots,N},~{x_i\in\{0,1\}}$ are available. These likelihoods are combined pair-wise by going through a data dependency graph (DDG) with $N\log N$ nodes, which are grouped into $\log N$ \emph{stages}. The output on the left side of the graph is $P(y_1^N,\hat{u}_1^{i-1}|u_i),~{u_i\in\{0,1\}},~{i = 1,\hdots,N}$. Hard decisions are taken according to
\begin{align}
	\hat{u}_i & = \left\{ \begin{matrix} \arg \max _{u_i \in \{0,1\}} P(y_1^N,\hat{u}_1^{i-1}|u_i), & i \in \mathcal{A}, \\ 0, & i \in \mathcal{A}^c.\end{matrix} \right. \label{eqn:ml}
\end{align}
The two pairs of incoming likelihoods at each node, denoted by $a_1^2$ and $b_1^2$, are combined in order to produce the \emph{intermediate likelihoods}, according to either $f: [0,1]^4 \rightarrow [0,1]^2$ or $g: [0,1]^4 \times \{0,1\} \rightarrow [0,1]^2$ with
\begin{align}
	f(a_1^2,b_1^2) & = \left(\frac{1}{2}\left(a_1b_1 + a_2b_2\right), \frac{1}{2}\left(a_2b_1 + a_1b_2\right)\right), \label{eqn:f}\\
	g(a_1^2,b_1^2, \hat{u}_s) & = \left(\frac{1}{2}a_{1+\hat{u}_s}b_1,\frac{1}{2}a_{2-\hat{u}_s}b_2\right), \label{eqn:g}
\end{align}
where $\hat{u}_s$ is called a \emph{partial sum}. Each partial sum is a linear combination of some of the previously decoded codeword bits \cite{Arikan2009}. The circle and square nodes of the DDG in Fig.~\ref{fig:sc} represent application of $f$ and $g$, respectively. If intermediate likelihoods are stored, then the computational complexity of SC decoding is $O(N\log N)$~\cite{Arikan2009}.

\begin{figure}
	\centering
	\includegraphics[width=0.4\textwidth]{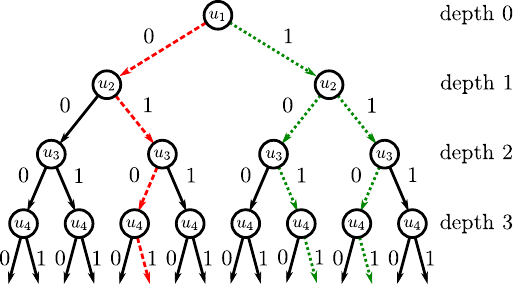}
	\caption{Decoding tree for $N=4$.}
	\label{fig:listsctree}
\end{figure}

\subsection{List SC Decoding}
Successive decoding can be described as a search procedure on a full binary tree. The $2^{(i-1)}$ nodes at depth $(i-1)$ represent $u_i$ given all possible choices for $\hat{u}_1^{i-1}$. The two outgoing edges of each node in the tree are labeled with the two possible choices for $\hat{u}_i$. A decoder explores one or more paths in the tree by deciding which edge to follow at each step based on some metric. The SC decoder explores a single path from the root to the leaves of the tree. It uses the likelihood in (\ref{eqn:ml}) as a metric for edges corresponding to non-frozen bits and it always follows the edge corresponding to $0$ for frozen bits. The SC decoder has the drawback that erroneous decisions at some point can never be recovered in the future. The list SC decoder, on the other hand, performs a breadth-first search on the tree under a complexity constraint. This constraint is enforced by discarding some of the paths at each step. Specifically, the list SC decoder with list size $L$ keeps track of $L$ paths simultaneously and also uses the likelihood in (\ref{eqn:ml}) as a path metric when encountering non-frozen bits. More formally, let $(\hat{u}_1^{i-1}(1), \hdots, \hat{u}_1^{i-1}(L))$ denote the $L$ distinct decoding paths after the $(i-1)$-th bit has been decoded. For every path $l \in \{1,\hdots,L\}$, there are two choices for $\hat{u}_{i}(l)$. Out of the resulting $2L$ paths, the $L$ paths with the highest metric are preserved. When bit $N$ is reached, the path with the highest metric is set as the decoded codeword. Decoding paths for an SC and a list SC decoder with $L=2$ are shown by the red dashed and green dotted lines in Fig.~\ref{fig:listsctree}, respectively.

\begin{figure}
	\centering
	\includegraphics[width=0.38\textwidth]{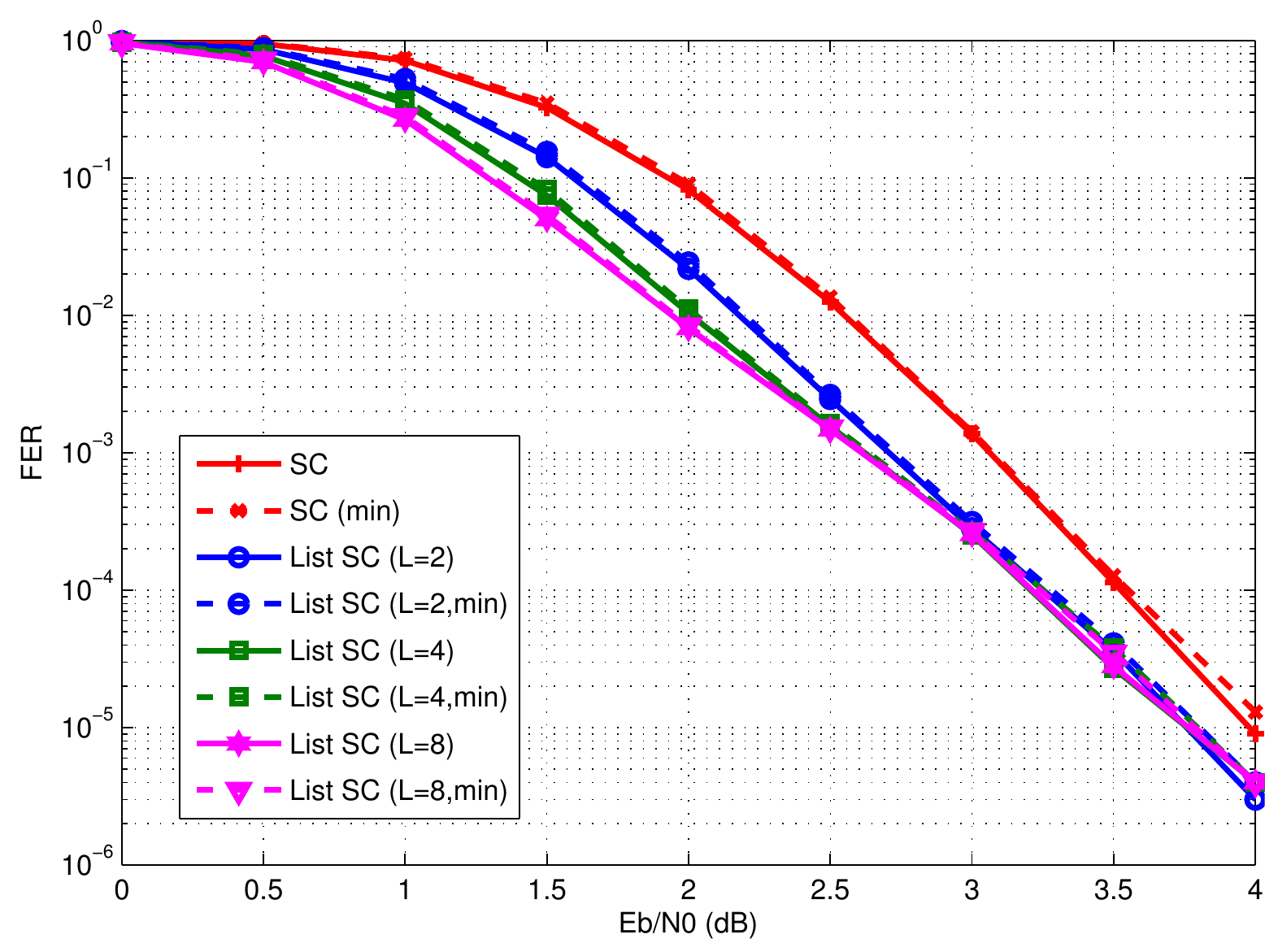}
	\caption{Frame error rate (FER) performance of a $(1024,512)$ polar code under SC and list SC decoding with various $L$.}
	\label{fig:listsc}
\end{figure}

All results in this paper are illustrated for $N = 1024$ and $R = 0.5$. The performance of list SC decoding over an AWGN channel for some practical list sizes is compared with the performance of SC decoding in Fig.~\ref{fig:listsc}. We performed $10^7$ Monte-Carlo simulations for each data point. We observe that the returns of increased list size are small for $L > 4$ and that at high SNR using $L >2$ provides almost no gain. However, at a FER of $10^{-2}$, which is a sensible target FER for many communications standards, e.g., \cite{IEEE802.11}, the gain of list SC decoding is not negligible.

\section{Algorithmic Considerations}\label{sec:alg}

For each path, the intermediate likelihoods, the partial sums, and the path itself are stored in memories. We call these three memories collectively the \emph{state-memories}. The content of each memory forms the \emph{state} of each path. After the path selection step, each of the initial $L$ paths is either discarded, kept, or duplicated, depending on whether it has zero, one, or two child nodes in the set of $L$ out of $2L$ largest metrics, respectively. In order to duplicate a path, in a straightforward implementation its state is copied from one state-memory to another state-memory, with some differences between the two copies that correspond to the two different choices for $\hat{u}_i$. It was shown in \cite{Tal2011} that list SC decoding can be performed with complexity $O(LN\log N)$ when using a \emph{lazy copy} technique. Our approach is to introduce an auxiliary \emph{pointer memory} in order to avoid the high complexity of likelihood copying.

The algorithm in Fig.~\ref{fig:alg1} describes list SC decoding. $LI$, $\hat{U}_s$, $\hat{u}$, and $p$ denote the likelihood, partial sum, path, and pointer memories, respectively. For simplicity we think of $LI$ as a three-dimensional memory which is indexed by the path index, the current stage index, and the bit index. Each element of $LI$ stores a \emph{likelihood pair}. The channel likelihood pairs are assumed to be stored in $LI(:,\log N, i),~i=1,\hdots,N,$ before \textsc{ListSC} is called. $\hat{U}_s$, $\hat{u}$, and $p$ are two-dimensional memories. Their first dimension is indexed by the path index and their second dimension is indexed by a combination of the partial sum, bit, and stage indices. The operation and structure of the pointer memory are described in the following section. The $2L$ path metrics are stored in the $L \times 2$ memory $M$. \textsc{PathSelection} takes $2L$ path metrics as input and outputs the indices of the parent paths corresponding to the paths with the $L$ best metrics, denoted by $l_p(l),~l=1,\hdots,L$, and the corresponding values for $\hat{u}(l,i)$. The straightforward copying approach is chosen for the partial sums and the paths because it can be carried out in a single clock cycle in hardware with small overhead due to the small size of the involved memories. Lines 4--10 of \text{ListSC} can be performed in parallel, since there are no data dependencies between the loop iterations. \textsc{UpdateStage}, which performs the likelihood updates for the given decoding stage using the update rules described in \cite{Arikan2009}, can also be executed in parallel for the $2^s$ nodes of the DDG that require updating at stage $s$~\cite{Leroux2013}.

\begin{figure}
\centering
\small
\algsetup{indent=1.5em}
\algsetup{linenosize=\small}
\begin{algorithmic}[1]
	\STATE \textbf{function} \textsc{ListSC}$(L)$
	\FOR{$i\leftarrow 1$ \TO $N$}
		\STATE $stage$ $\leftarrow$ index of first '1' in $\log N$-bit MSB-0 binary representation of $(i-1)$ (if $i=1$, then $stage$ $\leftarrow \log N$)
		\FOR{$l\leftarrow 1$ \TO $L$}
			\FOR{$s \leftarrow stage-1$ \TO $0$}
			\STATE $LI(l,s,:) \leftarrow \text{\textsc{UpdateStage}}(LI(p(l,s+1),s+1,:))$
			\STATE $p(l,s) \leftarrow l$
			\ENDFOR
			\STATE $M(l,:) \leftarrow LI(l,0,i)$
		\ENDFOR
		\IF{$i \in \mathcal{A}^c$ \AND $i < N$}
			\STATE $\hat{u}(l,i) \leftarrow 0,~l=1,\hdots,L$
		\ELSE
			\STATE $(l_p(:), \hat{u}(:,i)) \leftarrow \text{\textsc{PathSelection}}(M)$
			\STATE $p(l,:) \leftarrow p(l_p(l),:),~l=1,\hdots,L$
			\STATE $\hat{U}_s(l,:) \leftarrow \hat{U}_s(l_p(l),:),~l=1,\hdots,L$
			\STATE $\hat{u}(l,:) \leftarrow \hat{u}(l_p(l),:),~l=1,\hdots,L$
		\ENDIF
	\ENDFOR
	\RETURN $\hat{u}(1,:)$
\end{algorithmic}
\caption{\textsc{ListSC}: List SC decoding with list size $L$.}\label{fig:alg1}
\end{figure}

\subsection{Low-Complexity State Copying}\label{subsubsec:copying}
In this section, we describe the function of the pointer memory $p$. Assume that, for the code in Fig.~\ref{fig:sc}, we have $L=2$, and $u_1,u_2$ are non-frozen while $u_3,u_4$ are frozen. Decoding starts with one (empty) path. The path metrics for $\hat{u}_1 = 0,1,$ are calculated using the SC procedure based on the contents of the first state-memory. The intermediate likelihoods which are produced are written to the first state-memory. In general, the intermediate likelihoods which are produced for path $l \in \{1,\hdots,L\}$, are written to the $l$-th state-memory. Instead of taking a hard decision on $\hat{u}_1$ as the SC decoder would, the list SC decoder duplicates the (empty) parent path and extends the first copy with $\hat{u}_1 = 0$ and the second copy with $\hat{u}_1 = 1$. The SC procedure for the two new paths requires the intermediate likelihoods produced by their parent path in order to calculate the path metrics for $\hat{u}_2 = 0,1$. These likelihoods are located in the first state-memory, since they were produced by the first path in the previous decoding step. The intermediate likelihoods produced by the SC procedure for $\hat{u}_2 = 0,1,$ for the first and second paths are written to the first and second state-memory, respectively. From lines 5--8 of the list SC algorithm in Fig.~\ref{fig:alg1}, we see that the SC procedure does not process all stages of the decoding graph for each $u_i$. So, if after $\hat{u}_2$ has been processed the list SC decoder follows a new path whose parent is the second path, it has to read the intermediate likelihoods for the stages which were not processed when decoding $\hat{u}_2$ from the first state-memory, and the intermediate likelihoods for the stages which were processed when decoding $\hat{u}_2$ from the second state-memory. The auxiliary pointer memory $p$ of dimension $L \times (\log N-1)$ keeps track of which memory stores each path's likelihood for each stage. When a decision for $\hat{u}_2$ needs to be made, there are four candidate paths, out of which the two paths with the best metrics are kept, while the remaining two are discarded. Now, instead of copying the intermediate likelihoods of the paths that we want to keep, it suffices to copy the references to the state-memories contained in the pointer memory. Since $u_3$ and $u_4$ are frozen, both paths are extended with $\hat{u}_3 = 0$ and $\hat{u}_4 = 0$ and the best path is declared as the decoded codeword.


\begin{figure*}
\centering
\begin{minipage}{0.74\textwidth}
\subcaptionbox{List SC decoder architecture with details of the structure of the memory cells.}{\includegraphics[width=0.9\textwidth]{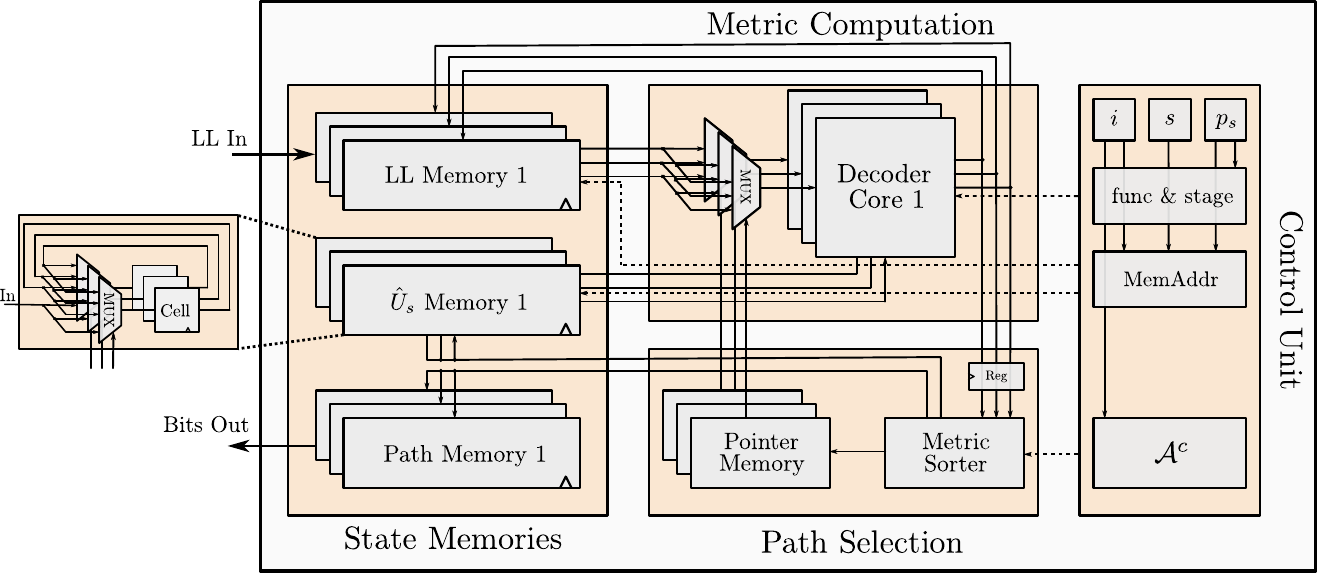}}
\end{minipage}%
\begin{minipage}{0.24\textwidth}
\subcaptionbox{Pointer memory (top) and metric sorter (bottom).}{\includegraphics[width=0.9\textwidth]{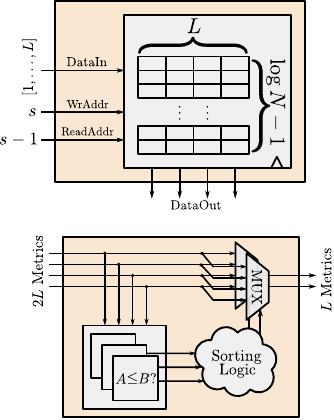}}
\end{minipage}
\caption{Block diagram of the proposed list SC decoder architecture.}\label{fig:archhigher}
\end{figure*}

\subsection{Likelihood Representation}
SC decoding can be carried out in the log-likelihood ratio (LLR) domain by modifying (\ref{eqn:f})--(\ref{eqn:g})~\cite{Leroux2011}. LLRs provide reduced storage requirements, increased numerical stability, as well as simplified computations with respect to a likelihood based implementation. The list SC decoding algorithm is described using likelihoods and log-likelihoods (LLs) in \cite{Tal2011} and \cite{Chen2013}, respectively. LLRs can be converted to LLs by using $\text{LL}(x_i) = \ln \frac{\text{exp}({\text{LLR})^{1-x_i}}}{1 + \text{exp}({\text{LLR}})}$. However, this conversion assumes that $\text{LL}(0)+\text{LL}(1)=1$, which is not true in general. Thus, each LL is normalized by a different factor, so the ordering of the path metrics will be affected and they can no longer be used to choose the $L$ best paths. For this reason, in our decoder the likelihoods are represented in LL form, which also simplifies the computations in (\ref{eqn:f})--(\ref{eqn:g}) and provides numerical stability, but requires more storage. We use negative LLs, which are always positive numbers and do not require a sign bit, to make the binary representation more compact. Assuming transmission over an AWGN channel with noise variance $\sigma ^2$, the negative LLs are
\begin{align}
	LL(x_i) = -\ln W(y_i|x_i)	& = \frac{(y_i-\mu(x_i))^2}{2\sigma ^2} + \ln \sqrt{2\pi \sigma ^2}, \label{eq:ll}
\end{align}
where $\mu (x_i) = 1- 2x_i$, $x_i \in \{0,1\},$ is the modulated version of codeword bit $x_i$. Using negative LLs, (\ref{eqn:f}) and (\ref{eqn:g}) become
\begin{align}
	f(a_1^2,b_1^2) 		& = (\text{min}^* (a_1+b_1,a_2+b_2), \nonumber \\
				& \qquad \qquad \quad \text{min}^* (a_2+b_1,a_1+b_2)), \label{eq:fll}\\
	g(a_1^2,b_1^2, \hat{u}_s) 	& = \left(a_{1+\hat{u}_s} + b_1,a_{2-\hat{u}_s} + b_2\right), \label{eq:gll}
\end{align}
where $\min ^* (a,b) = \min (a,b) + \ln \left(1+e^{-|a-b|}\right)$. The $f$ function is simplified by using an approximation that ignores the $\ln (\cdot)$ term. In Fig.~\ref{fig:listsc}, the performance of SC and list SC decoding with this approximation are plotted using dashed lines. There is practically no difference in performance with respect to the exact implementation for the used blocklength and list sizes. Our simulations show that the loss becomes slightly larger as the blocklength is increased. For example, for $N = 2^{15}$ the loss is approximately $0.1$~dB. Moreover, let $c,d > 0$ be constants. Then, for any $a,b \geq 0$, we have
\begin{align}
	\min (ca+d,cb+d) & = c \min (a,b)+d, \\
	(ca+d) + (cb+d)	& = c(a+b)+2d.
\end{align}
At each stage of SC decoding only one type of function is used, so the constant terms are common for all involved calculations. Thus, they can be recursively factored out and removed without affecting the ordering of the path metrics. So, we can use $LL(x_i) = (y_i-\mu(x_i))^2$, which is easier to handle by the quantization step.


\section{List SC Decoder Architecture}\label{sec:arch}

The list SC decoder is a combination of three components. The first component is the metric computation unit (MCU), which calculates the metrics for each path using the sequential SC procedure. The second component, called the \emph{state-memories} component, consists of $L$ state-memories, which the MCU uses to compute the $2L$ path metrics. Moreover, a third component manages the tree search by performing \emph{path selection} based on the metrics that are calculated by the MCU. An overview of the proposed list SC decoder architecture is presented in Fig.~\ref{fig:archhigher}(a). The MCU contains $L$ \emph{SC decoder cores}, which perform the metric calculation based on the state that they are supplied with. Multiplexers are responsible for redirecting the correct LLs to each decoder core, according to the entries of the pointer memory. The path selection unit contains a sorter which finds the $L$ best metrics out of $2L$ options, along with the path index and the value of $\hat{u}_i(l)$ from which they resulted, and the pointer memory, which manages the memory read access of the SC decoder cores.

\begin{figure}
	\centering
	\includegraphics[width=0.38\textwidth]{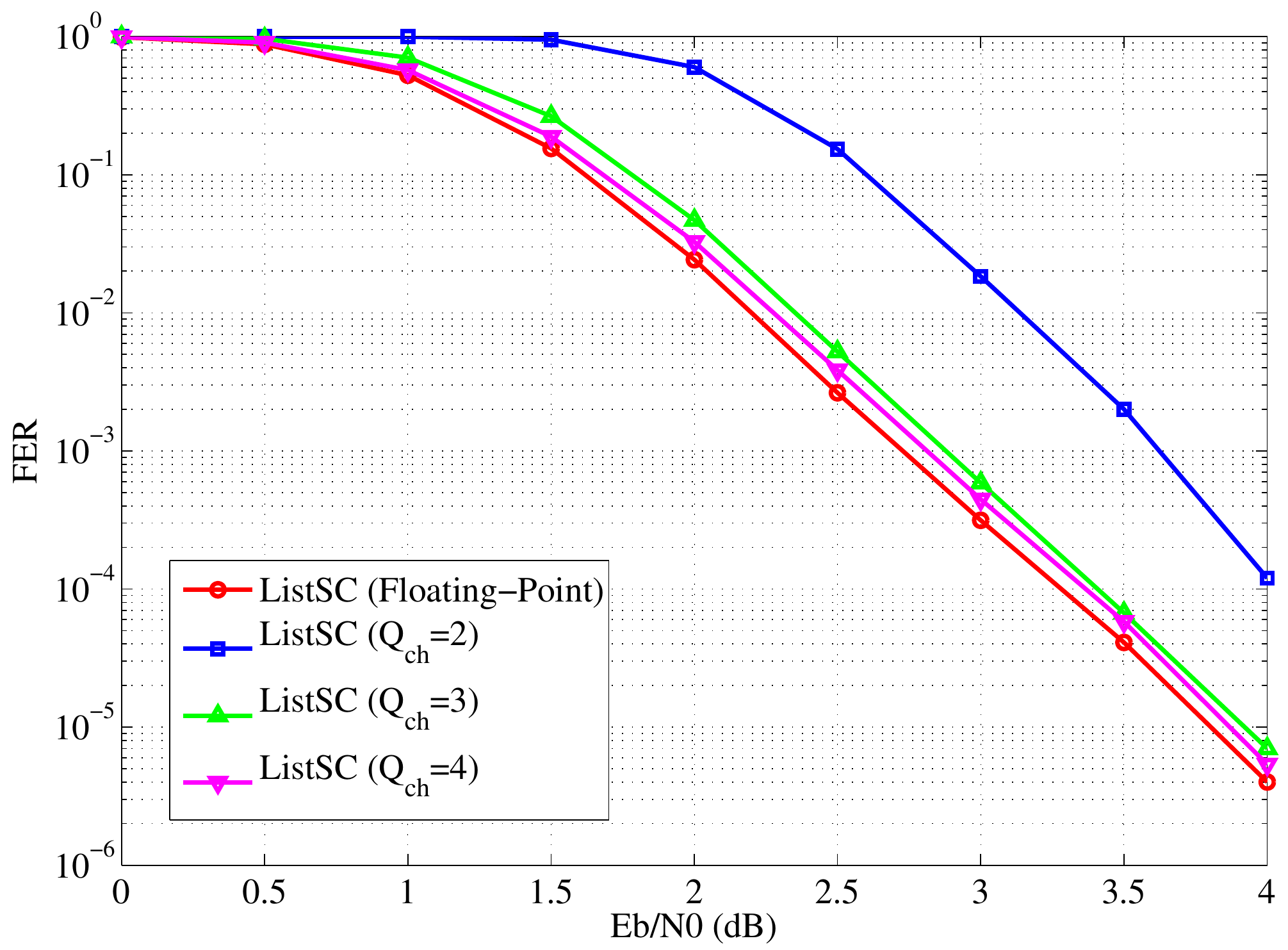}
	\caption{FER performance of a $(1024,512)$ polar code under floating-point and fixed-point list SC decoding for $L=2$.}
	\label{fig:listscquant}
\end{figure}


\subsection{LL Quantization}
Since the LLs are positive numbers, as SC decoding moves towards stage $0$, their dynamic range increases. When an LL pair saturates, it is useless for making a decision. Thus, when using LLs, it is crucial to avoid saturation. In (\ref{eq:fll}) and (\ref{eq:gll}), two numbers with the same dynamic range are added. The simplest way to avoid all saturations is to increase the number of bits used to store the LLs by one bit per stage. This way, the only performance degradation with respect to the floating point implementation comes from the quantization of the channel LLs. Let $Q_{ch}$ denote the number of bits used for the quantization of the channel LLs. The performance of the list SC decoder under various quantization bit-widths using a uniform quantizer with quantization step $\Delta = 1$ is presented in Fig.~\ref{fig:listscquant}. For the remainder of this paper, we choose $Q_{ch} = 3$, since the degradation with respect to the floating point and to the $Q_{ch} = 4$ implementations is very small.

\subsection{Metric Computation Unit}
The architecture of the SC decoder cores contained in the MCU is derived from the log-likelihood ratio (LLR) based architecture of \cite{Leroux2013}, which was modified to implement LL based SC decoding. Each decoder core consists of $P$ processing elements (PEs) that operate on up to $P$ nodes of each stage of the DDG in parallel. For fair comparison, we chose $P=64$ as in \cite{Mishra2012}. Three counters track the index $i$ of the bit that is currently being decoded, the current stage $s$ within the decoding graph, and the current part within the stage $p_s$ for the stages that require more than one cycle to be processed. All control signals and memory addresses are generated based on $(i,s,p_s)$. The maximum LL bit-width, denoted by $Q_{\max}$, determines the width of the PEs. Using the LL quantization scheme described previously, we have $Q_{\max} = Q_{ch} + \log N$. The PEs implement both (\ref{eq:fll}) and (\ref{eq:gll}). An additional input is used to choose between the $f$ and $g$ outputs. Due to the choice of quantization scheme, no overflow checks are needed. The MCU contains $L$ $L$-to-$1$ multiplexers, which are controlled by the pointer memory in the path selection unit and redirect the correct LLs to each SC decoder core.


\subsection{State-Memory}
SC decoding can be implemented by storing 2N LL pairs \cite{Leroux2013}, requiring a total of 4N data words. The $N$ first pairs that correspond to the channel LLs are never overwritten during SC decoding. Thus, only one copy of the channel LL memory is needed, from which all decoder cores can read. The remaining $N$ memory position pairs have to be distinct for each path. The number of required memory position pairs is $(L+1)N$ and the total number of bits used for LL storage is
\begin{align}
	B_{\text{LL}} 	& = 2 \left(NQ_{ch} + L \sum _{i=0}^{\log N-1}2^i\left( Q_{ch} + \log N - i \right)\right)\\
			& = (2L+2)NQ_{ch}+2L(2N - \log N - Q_{ch} - 2).
\end{align}
There are $L$ partial sum and $L$ path memories, with $N$ memory positions of $1$ bit each, resulting in a total of $2LN$ bits. The architecture of the partial sum memories is identical to the one used in \cite{Leroux2013}. In order to complete the state copying step in a single cycle, all the contents of each of the $L$ partial sum memories can be copied to and from one another by means of crossbars, as illustrated in Fig.~\ref{fig:archhigher}. The same holds for the path memories. So, the number of bits per MCU state is
\begin{align}
	B_{\text{tot}}	& = (2L+2)NQ_{ch}+2L(3N - \log N - Q_{ch} - 2).
\end{align}
\subsection{Path Selection}
\subsubsection{Metric Sorter}
For the path selection step, the $2L$ metrics are sorted in a single cycle. To minimize the delay, a radix-$2L$ sorter was implemented by extending the architecture presented in~\cite{Amaru2012} to support finding of the $L$ smallest values, instead of only the $2$ smallest values. This sorter requires $2L(2L-1)/2$ comparators of $Q_{\max}$-bit quantities. Since a single sorter is needed, minimizing its size is not critical. In fact, the metric sorter occupies only 0.1\% and 0.8\% of the total decoder area for $L=2$ and $L=4$, respectively. A register is added between the output of the MCU and the metric sorter in order to reduce the length of the critical path. 
Unfortunately, decoding can not proceed before the choice of paths is made, so an idle cycle is introduced every time the output of the metric sorter is needed. This happens $RN$ times per codeword. Thus, by modifying the expression found in \cite{Mishra2012}, the number of cycles required to decode one codeword is
\begin{align}
	C_{\text{list}} = (2+R)N + \frac{N}{P}\log \frac{N}{4P}.
\end{align}
If we ignore the second term, which is small, then the overhead with respect to the case where we do not add a register is approximately $RN$ cycles, or $\frac{RN}{2N} = 50R$ percent. Nevertheless, adding the register leads to a higher throughput due to a higher clock frequency. The architecture of the metric sorter is presented in Fig.~\ref{fig:archhigher}(b).
\subsubsection{Pointer Memory}
The pointer memory contains $L \times (\log N-1)$ elements. Each element can take on $L$ distinct values, so we need $\lceil\log L\rceil$ bits for the representation. In total, the pointer memory contains $L \lceil\log L\rceil (\log N-1)$ bits. For example, for $L=2,4$ and $N=1024$, this translates to $18$ and $72$ bits, respectively, which is negligible. This memory also has the copying functionality that the partial sum and path memories provide. The architecture of the pointer memory is presented in Fig.~\ref{fig:archhigher}(b).

\begin{table}
	\centering
	\caption{Synthesis results and comparison.}
	\label{tab:synthesis}
	\begin{threeparttable}
	\begin{tabular}{ccccc}
	\toprule
			& \multicolumn{2}{c}{Proposed Architecture}	& \cite{Leroux2013} & \cite{Mishra2012}\\
	\midrule
	Algorithm					& \multicolumn{2}{c}{List SC}			& SC 			& SC \\
	Code Length					& \multicolumn{2}{c}{$N = 1024$}		& $N = 1024$ 	& $N = 1024$ \\
	List Size					& $L=2$				& $L=4$ 			& n/a 			& n/a \\
	Cell Area					& 1.60~mm$^2$		& 3.53~mm$^2$ 		& 0.31~mm$^2$ 	& 1.71~mm$^2$\\
	Scaled to 65~nm				& 0.84~mm$^2$		& 1.85~mm$^2$ 		& 0.31~mm$^2$ 	& 0.22~mm$^2$ \\
	Clock Freq.					& 459~MHz\tnote{1}	& 314~MHz\tnote{1} 		& 500~MHz 	& 150~MHz \\
	Throughput					& 181~Mbps			& 124~Mbps 			& 246~Mbps 		& 49~Mbps \\
	Technology					& \multicolumn{2}{c}{UMC 90~nm}			& TSMC 65~nm 	& 180~nm \\
	\bottomrule
	\end{tabular}
	\begin{tablenotes}
		\item [1] We used the typical timing model at $25^\circ$C and $1$V supply voltage.
	\end{tablenotes}
	\end{threeparttable}
\end{table}

\section{Synthesis Results \& Conclusion}\label{sec:results}
Synthesis results for $N=1024$ and $L~=2,4$ using a UMC~90nm CMOS technology are shown in Table~\ref{tab:synthesis}. Since there exist no other list SC decoder architectures in the literature, we try to quantify the additional hardware complexity that is required to reap the decoding gain benefits of list SC decoding by comparing our design with the existing SC decoder synthesis results of~\cite{Leroux2013} and the chip results of~\cite{Mishra2012}.

In this work, the first list SC decoder architecture in the literature was presented. It was also described how to avoid copying of the intermediate likelihoods by copying between pointers instead of the actual values.

\section*{Acknowledgment}
The authors would like to thank Pascal Giard and Gabi Sarkis (McGill University), and Mani Bastani Parizi (EPFL) for helpful discussions. This project was kindly supported by the Swiss NSF under Project ID 200021\_149447.


\ifCLASSOPTIONcaptionsoff
  \newpage
\fi

\end{document}